\documentclass{ws-ijmpa}
\usepackage[super,compress]{cite}
\usepackage{graphicx}

\def\d{\textrm{d}}
\def\ds{\stackrel{\star}{,}}

\def\nn{\nonumber}
\def\be{\begin{equation}}             \def\ee{\end{equation}}
\def\ba#1{\begin{array}{#1}}          \def\ea{\end{array}}
\def\bea{\begin{eqnarray} }           \def\eea{\end{eqnarray} }
\def\beann{\begin{eqnarray*} }        \def\eeann{\end{eqnarray*} }
\def\beal{\begin{eqalign}}            \def\eeal{\end{eqalign}}
             
\def\bsubeq{\begin{subequations}}     \def\esubeq{\end{subequations}}
\def\bitem{\begin{itemize}}           \def\eitem{\end{itemize}}

\def\m{\mu}

\begin{document}
\markboth{M. D. \'Ciri\'c, N. Konjik, A. Samsarov}{ Duality between noncommutative and effective commutative description}

%
\catchline{}{}{}{}{}
%

\title{Noncommutative scalar field theory in a curved background: duality between noncommutative and effective commutative description }

\author{Marija Dimitrijevi\'c \'Ciri\'c }

\address{Faculty of Physics, University of Belgrade, Studentski trg 12\\
Belgrade, 11000, Serbia\\
dmarija@ipb.ac.rs}

\author{Nikola Konjik}

\address{Faculty of Physics, University of Belgrade, Studentski trg 12\\
Belgrade, 11000, Serbia\\
konjik@ipb.ac.rs}

\author{Andjelo Samsarov }

\address{Rudjer Bo\v skovi\'c Institute, Bijeni\v cka c.54, \\
HR-10002 Zagreb, Croatia\\
asamsarov@irb.hr}

\maketitle

\begin{history}
\received{Day Month Year}
\revised{Day Month Year}
\end{history}

\begin{abstract}
We study  a noncommutative (NC) deformation of  a charged scalar field,  minimally coupled
to a classical  (commutative) Reissner–Nordstr\" om-like
background. The deformation is performed
via a particularly chosen Killing twist to ensure that the geometry remains undeformed  (commutative). An action describing a NC scalar field minimally coupled to the RN geometry is manifestly invariant under the deformed $U(1)_\star$ gauge symmetry. We find the equation of motion and conclude that the same equation is obtained  from the commutative theory in a modified geometrical background described by an effective metric. This correspondence we call "duality between formal and effective approach".  We also show that a NC deformation via semi-Killing twist operator cannot be rewriten in terms of an effective metric. There is dual description for those particular deformations.

\end{abstract}

\ccode{PACS numbers:}

\newpage

\tableofcontents

\newpage

\section{Introduction}
Black holes (BH) can be perturbed  in  different ways. For example, a perturbation  may be  a
consequence of a gravitational collapse of matter to form a black hole, or it could happen simply as a result of   merging of two black  holes \cite{ligo}
that brought about the first detection of gravitational waves, or simply by infall of matter into a
supermassive black hole. In particular scenario with infalling of matter into a BH, it can possibly give us some specific information about quantum behaviour of matter and spacetime.

From many theoretical arguments, we expect that we need to deform the usual concept of  spacetime  at higher energies. In particular, it is well known that the quantum
theory and general relativity together lead to a noncommutative (NC) description of space-time, one between many others possibilities to deform spacetime.
\cite{ahluwalia,dop1,dop2}. In this paper we discus a particular type of NC spacetime, the one obtained from a deformation   by
a Drinfeld twist. We derive the differential calculus along the lines of \cite{PMKLWbook, PL09} and
we use the Seiberg-Witten map \cite{jssw, seiberg} to derive the NC gauge theory. The choice of
twist is non-unique and there are various examples in the literature. Our choice is twist 
 we call "angular twist" since one of the vector fields which define the twist, $x\partial_y-y\partial_x$, is the generator
of rotations around the $z$-axis. The angular twist  has additional nice properties
when applied to some special curved backgrounds as we will explain through the paper \cite{prvirad}.

In next section we will briefly introduce NC scalar field theory. The NC fields will be expanded via Seiberg-Witten map.
In the third section, particular choice of geometry is chosen. Namely, scalar field is coupled with Reissner-Nordstrom-like metric. In the fourth section, we found equvivalence between NC Reissner-Nordstrom-like metric and some commutative effective metric. In the last section, we discuss a  semi-Killing twist and  show  on a simple example that it is not possible to find solution for an effective metric.

\section{Noncommutative scalar field}

In order to construct a NC theory, we choose a Killing twist. The twist is given by
\begin{equation}\label{AngTwist0Phi}
\mathcal{F}=e^{-\frac{i}{2}\theta ^{AB}X_{A}\otimes X_{B}}=\bar{\mathrm{f}}^\alpha \otimes 
\bar{\mathrm{f}}_\alpha. 
\end{equation}
Here $\theta ^{AB}$ is a constant antisymmetric matrix
\begin{equation}
\theta^{AB} =\left( {
\begin{array}{cc}
0 & a \\ 
-a & 0
\end{array}
} \right) ,  \notag
\end{equation}
with an arbitrary constant $a$.
Indices $A,B=1,2$, while $X_{1}=\partial _t$,
$X_{2}= x\partial_y - y\partial_x=\partial_\varphi$ are commuting vector fields,
$[X_{1},X_{1}]=0$. The twist operator is used to deform (twist) the underlying symmetry algebra, in this case Poincare algebra. Additionally, it also  deforms the corresponding  commutative spacetime manifold. For example, the standard product of functions  to $\star$-product in following way:
\begin{eqnarray}
f\star g &=& (\bar{\mathrm{f}}^\alpha f)(\bar{\mathrm{f}}_\alpha g)= \mu \{ e^{\frac{ia}{2} (\partial_t\otimes
\partial_\varphi - \partial_\varphi\otimes \partial_t)}
f\otimes g \}\nn\\
&=& fg + \frac{ia}{2}
(\partial_t f(\partial_\varphi g) - \partial_t g(\partial_\varphi f)) + 
\mathcal{O}(a^2) . \label{fStarg}
\end{eqnarray}

In the similar way, the wedge product between forms is deformed to the $\star$-wedge product
\begin{equation}
\omega\wedge_\star\omega^{\prime }= \bar{\mathrm{f}}^\alpha(\omega) \wedge 
\bar{\mathrm{f}}_\alpha(\omega^{\prime }).  
\end{equation}

For more details about construction of angular deformation, see the \cite{prvirad}.
The formalism of NC field theories described above enables us to study the behaviour of a NC scalar field in a
gravitational field of the Reissner–Nordstr\" om like geometry described by
\begin{equation}
    ds^2=f(r)\,dt^2-\frac{1}{f(r)}\,dr^2-r^2\, d\Omega.
\end{equation}

In fact, we can construct a more general action which describes the NC $U(1)_\star$ gauge theory of a complex
charged scalar field on an arbitrary background. The only requirement is that vector fields 
$\partial_t$ and $\partial_\varphi$ which enter definition of the twist (\ref{AngTwist0Phi}) are the Killing vectors of the given background. In that case, the twist will not act on a geometry, that is on the metric tensor $g_{\mu\nu}$ and its functions.

Gauge field one form $\hat{A}=\hat{A}_\mu \star \d x^{\mu}$ is introduced to the theory via a minimal coupling. The relevant
action is
\begin{eqnarray}
S[\hat{\phi}, \hat{A}] &=& \int {\Big( \d {\hat{\phi}} - i\hat{A} \star {\hat{\phi}} \Big)}^+
\wedge_\star  *_H \Big( \d \hat{\phi} -i\hat{A} \star \hat{\phi} \Big) \nonumber \\
&& - \int \frac{\mu^2}{4!} \hat{\phi}^+\star \hat{\phi} \epsilon_{abcd}~ e^{a} \wedge_\star
e^b\wedge_\star e^c \wedge_\star e^d
- \frac{1}{4 q^2} \int (*_H \hat{F}) \wedge_\star \hat{F}. \label{NCActionGeometric}
\end{eqnarray}
The action describes the massive charged scalar field theory where the mass of the field  ${\hat{\phi}}$ is $\mu$ and its charge is $q$. The two-form
field-strength tensor is defined as
\begin{equation}
\hat{F} = \d \hat{A} - \hat{A} \wedge_\star
\hat{A} =\frac{1}{2} \hat{F}_{\m\nu}\star \d x^{\mu} \wedge_\star \d
x^{\nu}. \label{NCF} 
\end{equation}
To construct a massive theory geometrically in proper way, it is important to introduce  vierbein one-forms $e^a=e^a_\mu\star \d x^{\mu}$ and $g_{\mu\nu} = \eta_{ab}e_\mu^a\star
e_\nu^b$. In index notation, the action is of the form
\begin{eqnarray}
S[\hat{\phi}, \hat{A}] &=& S_\phi + S_A, \nn\\ 
S_\phi &=& \int \d ^4x \, \sqrt{-g}\star\Big( g^{\mu\nu}\star D_{\mu}\hat{\phi}^+ \star
D_{\nu}\hat{\phi} - \mu^2\hat{\phi}^+ \star\hat{\phi}\Big) , \label{SPhi}\\
S_A &=& -\frac{1}{4q^2} \int \d^4 x\,\sqrt{-g}\star g^{\alpha\beta}\star g^{\mu\nu}\star
\hat{F}_{\alpha\mu}\star \hat{F}_{\beta\nu} .\label{SA}
\end{eqnarray}
The scalar field $\hat{\phi}$ is a
complex charged
scalar field which transforms in the fundamental representation of NC $U(1)_\star$. Its covariant
derivative is
\begin{equation}
D_\mu\hat{\phi} = \partial_\mu\hat{\phi} - i \hat{A}_\mu\star \hat{\phi} \label{DPhi} . \nn
\end{equation}
 Equation (\ref{NCF}) defines the components of NC field-strength tensor in the following form
\begin{equation}
\hat{F}_{\mu\nu} = \partial_\mu \hat{A}_\nu - \partial_\nu \hat{A}_\mu -i [\hat{A}_\mu \ds \hat{A}_\nu]
.\label{F}
\end{equation}
Note that up to now, the geometry is an arbitrary. However, it is
important that its Killing vectors are used to define deformed products since only in that case
the action (\ref{SA}) has this simple form. Note also that $\star$-products in $\sqrt{-g}\star
g^{\alpha\beta}\star g^{\mu\nu}$ can all be removed since the twist does not affect the geometry. 
The actions (\ref{SPhi}) and (\ref{SA})  are invariant under the infinitesimal
$U(1)_\star$
gauge transformations
defined in the following way:
\begin{eqnarray}\label{ncgaugetrans}
\delta_\star \hat{\phi} &=& i\hat{\Lambda} \star \hat{\phi}, \nn\\
\delta_\star \hat{A}_\mu &=& \partial_\mu\hat{\Lambda} + i[\hat{\Lambda} \ds \hat{A}_\mu],
\label{NCGaugeTransf}\\
\delta_\star \hat{F}_{\mu\nu} &=& i[\hat{\Lambda} \ds \hat{F}_{\mu\nu}],\nn\\
\delta_\star g_{\mu\nu} &=& 0,\nn
\end{eqnarray}
whith the NC gauge parameter $\hat{\Lambda}$.

\subsection{Seiberg-Witten map}

NC gauge transformations (\ref{ncgaugetrans}) do not close in the corresponding Lie algebra, but only in the universal enveloping algebra. That leads to potentially infinitely many degrees of freedom. To solve this problem, we  use   the Seiberg-Witten (SW) map \cite{seiberg}. This map
enables to express NC variables as functions of the corresponding commutative variables and  deformation parameter $a$.

 Expansions for an arbitrary abelian
twist deformation are well known  \cite{PLSWGeneral}. In case of the deformation by the twist 
(\ref{AngTwist0Phi}), 
 up to first
order in the deformation parameter $a$  we obtain
\begin{eqnarray}
\hat{\phi} &=& \phi -\frac{1}{4}\theta^{\rho\sigma}A_\rho(\partial_\sigma\phi + D_\sigma
\phi), \label{HatPhi}\\
\hat{A}_\mu &=& A_\mu -\frac{1}{2}\theta^{\rho\sigma}A_\rho(\partial_\sigma A_{\mu} +
F_{\sigma\mu}), \label{HatA}\\
\hat{F}_{\mu\nu} &=& F_{\mu\nu} - \frac{1}{2}\theta^{\rho\sigma}A_{\rho}(\partial_\sigma F_{\mu\nu}
+ D_\sigma F_{\mu\nu})+\theta^{\rho\sigma}F_{\rho\mu}F_{\sigma\nu}. \label{HatFmunu}
\end{eqnarray}
The $U(1)$ covariant derivative of $\phi$ is defined as $D_\mu \phi = (\partial_\mu - i A_\mu)
\phi$, while $D_\sigma F_{\mu\nu} = \partial_\sigma F_{\mu\nu}$ in the case of $U(1)$ gauge
theory. It is important to note that the coupling constant
$q$
between fields $\phi$ and $A_\mu$, the charge of $\phi$, is included into $A_\mu$, namely $A_\mu =
qA_\mu$, compare also (\ref{SA}).


\subsection{Expanded actions and equtions of motion}

 After applying the SW map and  expanding $\star$-products up to first order in deformation parameter $a$, the action (\ref{SPhi}) wil have the following form
\begin{eqnarray}\label{13}
S &=& \int
\d^4x\sqrt{-g}\,
\Big( -\frac{1}{4q^2}g^{\mu\rho}g^{\nu\sigma}F_{\mu\nu}F_{\rho\sigma}
+ g^{\mu\nu}D_\mu\phi^+D_\nu\phi \nonumber\\
&&
+\frac{1}{8q^2}g^{\mu\rho}g^{\nu\sigma}\theta^{\alpha\beta}(F_{\alpha\beta}F_{\mu\nu}F_{\rho\sigma}
-4F_{\mu\alpha}F_{\nu\beta}F_{\rho\sigma})\label{SExp}\\
&& + \frac{\theta^{\alpha\beta}}{2}g^{\mu\nu}\big( -\frac{1}{2}D_\mu\phi^+F_{\alpha\beta}
D_\nu\phi
+(D_\mu\phi^+)F_{\alpha\nu}D_\beta\phi + (D_\beta\phi^+)F_{\alpha\mu}D_\nu\phi\big) \Big)
.\nn 
\end{eqnarray}

Now we vary the action (\ref{SExp}) to calculate the equations of motion. Varying the action
with respect to $\phi^+$ we obtain
\begin{eqnarray}\label{EoMPhi}
&&  g^{\mu \nu} \bigg( (\partial_{\mu} - iA_{\mu})D_{\nu} \phi - \Gamma^{\lambda}_{\mu
\nu}D_{\lambda} \phi \bigg)
-\frac{1}{4} \theta^{\alpha \beta} g^{\mu \nu} \bigg(  (\partial_{\mu} - iA_{\mu})(
F_{\alpha \beta}D_{\nu} \phi)
- \Gamma^{\lambda}_{\mu \nu} F_{\alpha \beta}  D_{\lambda} \phi   \nonumber\\
&& - 2  (\partial_{\mu} - iA_{\mu})( F_{\alpha \nu}D_{\beta} \phi)  + 2
\Gamma^{\lambda}_{\mu \nu} F_{\alpha \lambda}  D_{\beta} \phi  -2  (\partial_{\beta} -
iA_{\beta})( F_{\alpha \mu}D_{\nu} \phi) \bigg) = 0.  
\end{eqnarray}
Since the
backround metric $g_{\mu\nu}$ is not flat, the corresponding Christoffel symbols
$\Gamma^{\lambda}_{\mu \nu}$ appear in the equation of motion.

\section{Scalar field in the Reissner–Nordstr\" om-like background}

Finally, let us specify the gravitational background to be  a Reissner–Nordstr\" om-like background. It means that we have arbitrary spherically symmetric non-rotation geometry with only non-zero component of  spherically symmetric electromagnetic potential given by $A_0 (r)$.   

We write the RN-like metric tensor once again
\begin{equation}
g_{\mu\nu}=
\begin{bmatrix}
f & 0 & 0 & 0  \\
0 & -\frac{1}{f} & 0 & 0 \\
0 & 0 & -r^2 & 0  \\
0 & 0 & 0 & -r^2\sin^2\theta
\end{bmatrix} \label{RNmetric}
\end{equation}
with $f$ is an arbitrary function of $r$. The corresponding Christoffel symbols are:
\begin{eqnarray}
&& \Gamma^r_{tt} = \frac{f}{2}f' ,\quad \Gamma^r_{rr} =
-\frac{f'}{2f} = -\Gamma^t_{rt}, \nn\\
&& \Gamma^r_{\varphi\varphi} = -rf\sin^2\theta,\quad 
\Gamma^r_{\theta\theta} = -rf,\nn \\
&& \Gamma^\theta_{\varphi\varphi} = -\sin\theta\cos\theta,\quad
\Gamma^\varphi_{\theta\varphi} = \cot\theta, \quad \Gamma^r_{r\theta} = \Gamma^\varphi_{\varphi r}
=\frac{1}{r} ,\label{ChrSymb}
\end{eqnarray}
where $f' = \partial_r f$.
The
scalar potential
$A_0$, leads to the only non-zero component of the field strength tensor $F_{r0}=\partial_r A_0$.

The only non-zero components of the NC deformation parameter $\theta^{\alpha\beta}$ are
$\theta^{t\varphi}=
-\theta^{\varphi t}=a$. Inserting all these into (\ref{EoMPhi}) gives the following equation
\begin{eqnarray}\label{extendedKG}
&&
\Big( \frac{1}{f}\partial^2_t -\Delta + (1-f)\partial_r^2 
-(f'+\frac{2f}{r})\partial_r + 2i\frac{A_0}{f}\partial_t -\frac{A_0^2}{f}\Big)\phi
\nonumber\\
&& -\frac{aF_{0r}}{r}
\Big(\frac{rf'}{2}\partial_\varphi
+ rf\partial_r\partial_\varphi \Big) \phi =0 ,\label{EomPhiExp1}
\end{eqnarray}
  with the usual Laplace operator $\Delta$. 

\section{Effective metric from first order noncommutative duality}

Now we show that equation (\ref{extendedKG}) can be obtained from a commutative theory in a modified geometry up to
the first order in the deformation parameter $a$. The metric tensor describing this modified geometry  we  call  
 effective metric. 

Equation (\ref{extendedKG}) can be written  symbolically as a commutative equation with corrections ${\mathcal{O}}(a)$ as 
\begin{equation} \label{KG1}
 \big( {\Box_{g}} + {\mathcal{O}}(a) \big) \Phi \equiv \bigg( g^{\mu \nu} \big( \nabla_{\mu} - i A_{\mu}  \big )  \big( \nabla_{\nu} - i A_{\nu}  \big) + {\mathcal{O}}(a) \bigg) \Phi= 0.
\end{equation}
 The operator  $\nabla_{\mu}$ is a covariant derivative with respect to the metric $g_{\mu \nu}$\footnote{$\nabla_{\mu} A^{\nu} = \partial_{\mu} A^{\nu} + \Gamma^{\nu}_{~\lambda \mu} A^{\lambda}$  and  $\nabla_{\mu} A_{\nu} = \partial_{\mu} A_{\nu} - \Gamma^{\lambda}_{~ \mu \nu} A_{\lambda}$.} and  $\Box_{g}$ is the Klein-Gordon operator for the metric $ g_{\mu \nu}$.  In the commutative limit  $a \rightarrow 0,$ all NC corrections and the KG equation reduce
to
\begin{equation} \label{KG2}
{\Box_{g'}}  \Phi \equiv  g'^{\mu \nu} \big( \nabla'_{\mu} - i A_{\mu}  \big )  \big( \nabla'_{\nu} - i A_{\nu}  \big)   \Phi = \frac{1}{\sqrt{-g'}} (\partial_{\mu} -iA_{\mu}) \bigg( \sqrt{-g'} ~ g'^{\mu \nu} \big( \partial_{\nu} - i A_{\nu} ) \bigg)  \Phi = 0.
\end{equation}
The term linear in  $a$ in (\ref{extendedKG}) could be combined with the operator  ${\Box_{g}}$ to yield the  operator ${\Box_{g'}}$ with a redefined metric that  absorbed the  noncommutative contributions in (\ref{extendedKG}). 
 More concisely, the question to be posed is if there exists a metric which is able to meet the requirement
\begin{equation} \label{KG3}
\begin{split}
 \big(  {\Box_{g}} + {\mathcal{O}}(a) \big) \Phi &\equiv \bigg( g^{\mu \nu} \big( \nabla_{\mu} - i A_{\mu}  \big )  \big( \nabla_{\nu} - i A_{\nu}  \big) + {\mathcal{O}}(a) \bigg) \Phi
\\
&= {\Box_{g'}}  \Phi  =   g'^{\mu \nu} \big( \nabla'_{\mu} - i A_{\mu}  \big )  \big( \nabla'_{\nu} - i A_{\nu}  \big)  \Phi\\
    &= \frac{1}{\sqrt{-g'}}
   (\partial_{\mu} -iA_{\mu}) \bigg( \sqrt{-g'} ~ g'^{\mu \nu} \big( \partial_{\nu} - i A_{\nu}  \big)  \bigg)  \Phi = 0,
\end{split}
\end{equation}
where $\nabla'_{\mu}$ is  a covariant derivative with respect to the new, effective metric $g'_{\mu \nu}$. We point out that the gauge potential did not change upon switching to a new setting  and rewriting dynamics of the system in terms of the effective metric.

In order to find the metric tensor which satisfies the requirement (\ref{KG3}), one may try  with the following ansatz
\begin{equation}\label{effectivemetric}
g_{\mu\nu}=\begin{pmatrix}
 f &0& 0 & 0\\
 0&-\frac{1}{f} &0 & g_{r \phi}\\
 0 &0& - r^2 & 0 \\
  0 & g_{r \phi}& 0 & - r^2 \sin \theta \\
\end{pmatrix},
\end{equation}
The new metric term   $g_{r \phi}$ is assumed to depend only on variables $r$ and $\theta,$ since we expect that $\partial_{t}$ and $\partial_{\phi} $ are Killing vectors for the  effective metric as well.  Moreover, it is assumed to be at least linear in NC parameter $a,$  $g_{r \phi} \sim {\mathcal{O}}(a)$ since the effective metric $g_{\mu \nu}$ has to reduce to the original RN metric  $g_{\mu \nu}$ in the limiting case $a \rightarrow 0$. The reason why we have chosen this ansatz is that we have only terms proportional to $\partial_r\partial_\varphi$ and $\partial_\varphi$ and this type of ansatz will give also these terms. The inverse of the metric tensor (\ref{effectivemetric}) is given by

\begin{equation}\label{inverseeffectivemetric}
g^{\mu\nu}=\begin{pmatrix}
 \frac{1}{f} &0& 0 & 0\\
 0&-f +\frac{f^2 g^2_{r \phi}}{f g^2_{r \phi} -  r^2 \sin^2 \theta} &0 & \frac{f g_{r \phi}}{f g^2_{r \phi} -  r^2 \sin^2 \theta}\\
 0 &0& - \frac{1}{ r^2 } & 0 \\
  0 &  \frac{f g_{r \phi}}{f g^2_{r \phi} -  r^2 \sin^2 \theta}  & 0 &  \frac{1}{f g^2_{r \phi} -  r^2 \sin^2 \theta} \\
\end{pmatrix},
\end{equation}
It can be seen that while off-diagonal elements have a leading correction term that is linear in $a$, the diagonal elements $g^{rr}$ and $g^{\phi \phi}$ have a leading correction term that is quadratic in $a$. 
Likewise, the determinant and the square-root of the determinant of the effective metric (\ref{effectivemetric}) have a leading correction term that is quadratic in the NC parameter $a,  ~~ \sqrt{-g} = r^2 \sin \theta + {\mathcal{O}}(a^2)$. These observations will have a crucial role in the subsequent analysis. 

The form of the metric (\ref{effectivemetric}) dictates which terms are going to survive after the equation 
(\ref{KG3}) is written out explicitly

\begin{eqnarray}
{\Box_{g}}  \Phi  &=& \frac{1}{\sqrt{-g}}
(\partial_{\mu} -iA_{\mu}) \bigg( \sqrt{-g} ~ g^{\mu \nu} \big( \partial_{\nu} - i A_{\nu}  \bigg)  \Phi \nn\\
&=& \frac{1}{\sqrt{-g}} \Bigg[ (\partial_t - i A_t)\bigg( \sqrt{-g}g^{tt} \big( \partial_t - i A_t   \big)  \bigg) + 
   (\partial_r - i A_r)\bigg( \sqrt{-g}g^{rr} \big( \partial_r - i A_r   \big)  \bigg) \nn\\
&& +  (\partial_r - i A_r)\bigg( \sqrt{-g}g^{r \phi} \big( \partial_{\phi} - i A_{\phi}   \big)  \bigg)
+ (\partial_{\theta} - i A_{\theta})\bigg( \sqrt{-g}g^{\theta \theta} \big( \partial_{\theta} - i A_{\theta}   \big)  \bigg) \nn\\
&& +  (\partial_{\phi} - i A_{\phi})\bigg( \sqrt{-g}g^{\phi r} \big( \partial_r - i A_r  \big)  \bigg)
+  (\partial_{\phi} - i A_{\phi})\bigg( \sqrt{-g}g^{\phi \phi} \big( \partial_{\phi} - i A_{\phi}  \big)  \bigg)  \Bigg] \Phi .\nn
 \end{eqnarray}
Taking into account the fact that the gauge potential has only time component, one finds that the equation of motion (\ref{KG3}) further  boils down to
\begin{eqnarray}
 \frac{1}{f} \bigg[ \partial_t^2 \Phi -2i A_t \partial_t  \Phi  \bigg] &+& \frac{1}{\sqrt{-g}} \bigg[ \partial_r \big( \sqrt{-g}g^{rr}   \big)  \bigg] \partial_r \Phi
+ g^{rr} \partial^2_r \Phi + \frac{1}{\sqrt{-g}} \bigg[ \partial_r \big( \sqrt{-g}g^{r \phi}   \big)  \bigg] \partial_{\phi} \Phi \nn\\
&+&  2g^{r \phi} \partial_r \partial_{\phi} \Phi + \frac{1}{\sqrt{-g}} \bigg[ \partial_{\theta} \big( \sqrt{-g}g^{\theta \theta}   \big)  \bigg] \partial_{\theta} \Phi + g^{\theta \theta} \partial^2_{\theta} \Phi  + g^{\phi \phi}  \partial^2_{\phi} \Phi = 0. \nn
\end{eqnarray}
Focusing only on terms in the above equation that are at most linear in $a$, and comparing them with  the equation  (\ref{extendedKG}) leads to the following two relations:
\begin{equation} \label{KGconditions}
\begin{split}
  -\frac{aF_{0r}}{2} f' \partial_{\phi} \Phi  &=  \frac{1}{\sqrt{-g}} \bigg[ \partial_r \big( \sqrt{-g}g^{r \phi}   \big)  \bigg] \partial_{\phi} \Phi,
 \\
 -aF_{0r} f  \partial_r  \partial_{\phi} \Phi &=  2 g^{r \phi}   \partial_r \partial_{\phi} \Phi.
\end{split}
\end{equation}
The solution to this set of relations, which is consistent with the requirement $g^{r \phi} =\frac{f g_{r \phi}}{f g^2_{r \phi} -  r^2 \sin^2 \theta}, $ finally gives for the dual effective metric
\begin{equation} \label{metric}
  g_{\mu \nu} =
\left( \begin{array}{ccccc}
  f & 0  & 0  & 0 \\
   0   & -\frac{1}{f} & 0 & \frac{ar^2F_{0r}}{2} \sin^2 \theta \\ 
   0  & 0 & -r^2  &  0 \\
   0  & \frac{ar^2F_{0r}}{2}\sin^2 \theta & 0 & -r^2 \sin^2 \theta \\
\end{array} \right)  
\end{equation}
and for its inverse metric
\begin{equation}\label{InvMetric}
g^{\mu \nu} =
\left( \begin{array}{ccccc}
  \frac{1}{f} & 0  & 0 & 0  \\
   0   & -f  &  0  &  -\frac{aF_{0r}}{2} f    \\ 
   0  & 0 & -\frac{1}{r^2 } & 0   \\
   0   & -\frac{aF_{0r}}{2} f  &  0 &  -\frac{1}{r^2  \sin^2 \theta } \\
\end{array} \right).
\end{equation}
Note that we demand $g_{\mu \nu} g^{\nu \rho} =  \delta_{\mu}^{~~\rho} + {\mathcal{O}}(a^2).$

We have thus shown that the equation of motion for a charged NC scalar field in a classical RN-like background, coupled  to  NC $U(1)$ gauge field
may be rewritten in terms of the equation of motion governing behaviour of a charged commutative scalar field (having the same charge $q$ as its NC counterpart), propagating in a modified RN-like geometry
\begin{equation} 
  {\rm d}s^2 = \Big(1-\frac{2MG}{r}+\frac{Q^2G}{r^2} \Big) {\rm d}t^2 - \frac{{\rm
d}r^2}{1-\frac{2MG}{r}+\frac{Q^2G}{r^2}} - aqQ \sin^2 \theta {\rm d} r {\rm d} \phi - r^2({\rm d}\theta^2 + \sin^2\theta{\rm d}\phi^2 ).
\end{equation}
We have seen that in the case when  we have an arbitrary spherically symmetric non-rotating geometry with only zero component of  spherically symmetric electromagnetic potential $A_0$, and when vector fields which define the twist operator are Killing vectors for the given geometry, it is possible to find a solution for the effective metric.

\section{Comments on more general noncommutative deformations}
In this section, we  briefly discuss a  semi-Killing twist and  show  on a simple example that it is not possible to find solution for an effective metric.

Let us define one particular semi Killing twist operator as
\begin{equation}
\mathcal{F} =  e^{-\frac{ia}{2} (\partial_t\otimes\partial_r - \partial_r\otimes\partial_t)} \nn
\end{equation}
for the RN-like metric. Following the steps described in Section 2, we find the Seiberg-Witten expanded (up to first order) action for the NC scalar field to be of the form
\begin{eqnarray}
S &=& \int
\d^4x \Big( \sqrt{-g}\, \big( g^{\mu\nu}D_\mu\phi^+D_\nu\phi -\mu^2\phi^+\phi +\frac{\mu^2}{2}\theta^{\alpha\beta}F_{\alpha\beta}\phi^+\phi \nonumber\\
&& + \frac{1}{2}\theta^{\alpha\beta}g^{\mu\nu}\big( -\frac{1}{2}D_\mu\phi^+F_{\alpha\beta}
D_\nu\phi +(D_\mu\phi^+)F_{\alpha\nu}D_\beta\phi + (D_\beta\phi^+)F_{\alpha\mu}D_\nu\phi\big) \big)\nn\\
&&-\frac{i}{2}\theta^{\alpha\beta}\partial_\alpha \big(\sqrt{-g}g^{\mu\nu}\big) D_\mu\phi^+D_\beta D_\nu\phi \big) \Big)
.\nn 
\end{eqnarray} 
This equation we compare with equation (\ref{13}).
We see that there is an the additional term $-\frac{i}{2}\theta^{\alpha\beta}\partial_\alpha \big(\sqrt{-g}g^{\mu\nu}) D_\mu\phi^+ D_\beta D_\nu\phi$. When we use the variational principle to calculate the equation of motion for the NC scalar field $\phi$, this new term  will lead to an equation that is third order in derivatives of the field $\phi$. This immediately signals that equation cannot be reduced to an equation of motion for a commutative scalar filed in an effective metric, since that equation is necessarily a second order differential equation.

Similar analysis can be applied to a NC scalar field theory defined by a twist operator that is neither Killing nor semi-Killing. The result will be similar to the result found here for the semi-Killing twist. Therefore, we calculate that the (first order) duality between a NC field theory and a commutative field theory on a modified background (geometry) only holds for particular types of NC deformations given by Killing twist operators.

\section*{Acknowledgments}

N.K. thanks the organisers of "SEENET-MTP Assesment Meeting and Workshop on Theoretical and Mathematical Physics 2022"  for the invitation to give a talk.  This research was supported by the Croatian Science Foundation Project No. IP-2020-02-9614
Search for Quantum spacetime in Black Hole QNM spectrum and Gamma Ray Bursts. The work of M.D.C. and
N.K. is supported by project 451-03-47/2023-01/200162 of the Serbian Ministry of Education and Science. This work
is partially supported by ICTP-SEENET-MTP Project NT-03 ”Cosmology-Classical and Quantum Challenges” in
frame of the Southeastern European Network in Theoretical and Mathematical Physics and the COST action CA18108
Quantum gravity phenomenology in the multimessenger approach.


\end{document}